\documentclass[conference]{IEEEtran}
\usepackage{footnote}
 \usepackage[pscoord]{eso-pic}
\newcommand{\placetextbox}[3]{
 \setbox0=\hbox{#3}
 \AddToShipoutPictureFG*{ \put(\LenToUnit{#1\paperwidth},\LenToUnit{#2\paperheight}){\vtop{{\null}\makebox[0pt][c]{#3}}}
 }
 }
 \placetextbox{.50}{0.055}{\textcolor{gray}{\small{This work is supported by Dell EMC who provided the hardware.  Any opinions from the authors do not reflect the views of the company.}}}

\ifCLASSINFOpdf
\else
\fi

\begin{document}
%
\title{Big Memory Servers and Modern Approaches to Disk-Based Computation}


\author{
\IEEEauthorblockN{Po Hao Chen}
\IEEEauthorblockA{Boston University\\
Boston, MA\\
bupochen@bu.edu}
\and
\IEEEauthorblockN{Kurt Keville}
\IEEEauthorblockA{Massachusetts Institute of Technology\\
Cambridge, MA\\
klk@mit.edu}

}


%


\maketitle

\begin{abstract}
The Big Memory solution is a new computing paradigm facilitated by commodity server platforms that are available today. It exposes a large RAM subsystem to the Operating System and therefore affords application programmers a number of previously unavailable options for data management. Additionally, certain vendor-specific solutions offer additional memory management options that will result in better data reliability and access speeds. \end{abstract}


%
\IEEEpeerreviewmaketitle

\section{Introduction}
\IEEEPARstart{I}{n} recent years, datasets of interest are growing larger in size and at a faster rate.
Although we expect the capability of our processors to increase every few years, traditional memory models cannot keep up with that growth. While volatile memory such as DRAM is fast, it is limited by its capacity and cost. Large data must be placed in stable storage and retrieved when needed. Bottlenecks arise in performance-critical workloads, and therefore a new paradigm is required to handle emergence of this trend. This paper is targeted at improving High Performance Computing (HPC) and data-center applications in terms of scaling systems in affordable ways. 

A Big Memory server is one with a RAM subsystem ranging from 6-24 TB, larger than that available on contemporary systems. This is a somewhat arbitrary definition, but consider that it is difficult to get even 4TB of DRAM in a modern server motherboard.
Additionally, we look at the whole memory stack when discussing main memory. While we may never be able to do much off-processor improvements with L1 and L2 cache, we can likely do a lot with L3 (and L4).
In modern HPC centers, we often discuss the server architectures as multicore and multi-processor.  Our Big Memory techniques unequally favor multicore scale-up design improvements, since, in the perennial search to remove every source of latency in our compute stack, we remove network cards where we can.

Jun \textit{et al.} \cite{10.1145/2749469.2750412} showed  experimental evidence that, if servers executing a distributed computation have to retrieve data from disk even five percent of the time, their performance falls to a level that’s comparable with flash. Big Memory systems address this issue and deliver improvements to a range of applications. 

Many performance-critical workloads can benefit from such systems. Examples may include real-time analytic, financial data analysis, and other Machine Learning applications. For instance, recommendation engine models can be hundreds of gigabytes in size and their embedding table can be even terabytes. It is important to consider scalability in production, and the model inference time can be greatly reduced if the computation can be done in-memory.

Additionally, a large RAM subsystem can be used to enumerate the possible branches on a computation tree and is useful for implementing large proof-by-exhaustion of a theorem. For example, mathematicians spent over two decades hunting the number of moves an optimal algorithm would take in the worst case to solve any configuration of the Rubik's Cube. Before cracking the mystery, its upper-bound was reduced to 26 by the use of disk-based computation \cite{article}. In this work, Kunkle and Cooperman enumerated and stored 4.3 x $10^{19}$ possible configurations of the cube which required terabytes of main memory. They argued that because distributed disks offer approximately the same bandwidth as a single RAM subsystem, the local disks of a computer cluster can be regarded as a single large RAM subsystem.

In Section II, we explore the possible existing solutions to handle the situations when the size of data becomes greater than memory. In Section III, we investigate performance of a disk-based computation alternative using NVMe NAND-flash SSDs. In Section IV, we take a look at the future directions of the industry to serve the Big Memory community.

\section{The Incumbents \& Existing Solutions}
\subsection{Virtualization Technologies}
Historically, the workaround to easing memory-bound problem restrictions is through the employment of virtualization strategies. In HPC, these have been in place for quite a few years from companies such as ScaleMP \cite{ScaleMP}. They provide a scale-up solution by aggregating multiple independent systems over fast interconnects to create a single virtual cluster with large shared memory.
 
\subsection{Intel Optane and MemVerge}
Intel Optane Memory \cite{Intel} has very low latency and is non-volatile, making it a good secondary memory device to complement DRAM. MemVerge \cite{memverge} has partnered with Intel to create a memory abstraction layer known as ``Software Defined Memory" by taking advantage of virtualization in KVM to create a Big Memory solution that outperforms DRAM. However, it is currently implemented only on an enterprise level due to its high cost. 
\section{NVMe-centric Approach }
Flash storage is an order of magnitude less expensive compared to DRAM in terms of costs and power consumption.
We propose a software-defined memory abstraction layer with relatively cheaper NVMe NAND-flash SSDs to lower the total cost of ownership of HPC centers without sacrificing the performance drastically.
\subsection{Testbed}
As a proof of concept, we used a PowerEdge R7525 equipped with AMD EPYC 7763 64-Core Processors, 512 GB of DDR4-3200 RAM, with 16 3.2 TB PCIe 4.0 NVMe SSDs to measure the possible throughput for an NVMe-based Big Memory server. We ran two types of tests: a sequential test with a single thread and queue depth of 32 and a random test with 8 threads and queue depth of 8. We obtained the throughput numbers described in Table I while DDR4-3200 throughput is approximately 25000 MB/s.
\begin{table}[ht]
    \label{table:throughputnvme}
    \centering
    \begin{tabular}{| c|c|c|}
    \hline
    Tests /  Num of drives & 8 & 16\\
    \hline
    Sequential Q32T1   Read & 20722MB/s & 20805MB/s  \\
    Sequential Q32T1 Write & 11888MB/s & 19204MB/s \\
    \hline
    Random 4KB Q8T8 Read & 3777MB/s & 3972MB/s \\
    Random 4KB Q8T8 Write & 1445MB/s & 1923MB/s \\  
    \hline
    \end{tabular}
    \vspace{2mm} 
    \caption{Throughput for NVMe SSD (RAID 0)}
\end{table}
\subsection{Challenges}
Data is accessed in pages in flash as opposed to the more granular access pattern DRAM has. Many algorithms require fine-grained access and the performance may drop because the fetched page is not completely utilized as shown in Table I. Therefore, a transformation of random read/write to sequential ones must be implemented.
Moreover, NAND flash has a limited lifespan. The mechanics of flash storage traps electrons inside insulating materials and discerns ones and zeros by detecting the electric fields inside the memory cells. Occasionally, electrons can be permanently trapped due to quantum tunneling and the accumulation eventually creates a field that compromises the ability of a memory cell to distinguish the bits.

To avoid wearing out the SSDs, we recommend using only 25\% of each drive and distribute writes across multiple flash devices. This technique is known as overprovisioning; it greatly increases wear-leveling and performance. Our server has approximately 50TB of flash storage, but it is unlikely we need 50TB of working memory in contemporary workloads. If we do, we can increase the number of drives and resultant throughput. QTY 16 3.2 TB PCI-e 4.0 (x 4) drives only use 96 lanes of our 128 available.
\section{Future Directions}
\subsection{PCI-e Gen 5}
Table II shows the throughput performance of PCI-e links double every generation. The future PCI-e bandwidth will allow better and faster SSDs to match the speed of DDR modules (rows correspond to generations, columns to number of lanes).
\begin{table}[ht]
    \centering
    \begin{tabular}{| c|c|c|c|}
     \hline
     & $\times$4 & $\times$8 &
    $\times$16\\
    \hline
    1.0 &  1.000 GB/s & 2.000 GB/s & 4.000 GB/s \\
    2.0 & 2.000 GB/s & 4.000 GB/s & 8.000 GB/s\\
    3.0 &  3.938 GB/s & 7.877 GB/s & 15.754 GB/s \\
    4.0 &  7.877 GB/s & 15.754 GB/s & 31.508 GB/s\\
    5.0 &  15.754 GB/s & 31.508 GB/s & 63.015 GB/s  \\
    \hline
    \end{tabular}
    \vspace{2mm} 
    \caption{PCI Express Throughput Performance}
   
\end{table}
\subsection{Compute Express Link™ (CXL)}
Compute Express Link™ (CXL™) is an industry-supported Cache-Coherent Interconnect for Processors, Memory Expansion and Accelerators \cite{das}. CXL technology maintains memory coherency between the CPU memory space and memory on attached devices, which allows resource sharing for higher performance, reduced software stack complexity, and lower overall system cost. This permits users to simply focus on target workloads as opposed to the redundant memory management hardware in their accelerators.

CXL is designed to be an industry open standard interface for high-speed communications, as accelerators are increasingly used to complement CPUs in support of emerging applications such as Artificial Intelligence and Machine Learning.

\section{Conclusion}
We espouse a Big Memory approach that  leverages a processor agnostic approach like CXL that will be able to take advantage of NVMe-centric memory support. We anticipate that the industry will move into a more tiered approach to a multilevel memory model as flash costs decrease and PCI-e bus speeds increase to make them competitive with traditional DRAM solutions.

\section*{Acknowledgement}
We thank the MIT Lincoln Laboratory Supercomputing Center team for providing reviews and edit suggestions.



%
\bibliography{biblo.bib}

\end{document}